\documentclass[prl,aps,twocolumn,showpacs,showkeys,superscriptaddress,groupedaddress]{revtex4}  
\usepackage{graphicx}  
\usepackage{dcolumn}   
\usepackage{bm}        
\usepackage{amssymb}   
\usepackage{color}

\begin{document}
\title{The strangest non-strange meson is not so strange after all}

\author{S.~Ceci}
\email{sasa.ceci@irb.hr}
\affiliation{Rudjer Bo\v{s}kovi\'{c} Institute, Bijeni\v{c}ka  54, HR-10000 Zagreb, Croatia}
\author{M.~Vuk\v si\' c}
\affiliation{University of Oxford, Oxford OX1 2JD, United Kingdom}
\author{B.~Zauner}
\affiliation{Institute for Medical Research and Occupational Health, Ksaverska 2, HR-10000 Zagreb, Croatia}


\begin{abstract}
The broad non-strange meson $f_0$(500) is considered to be a non-Breit-Wigner resonance. It is generally assumed that if a resonance is broad it can hardly be described by a Breit-Wigner form. We show here that is not true neither for the broad Z boson, nor for the much narrower $\Delta(1232)$, and explain why the $f_0$(500) fits with them perfectly. Key differences between them boil down to a single parameter: the angle at which we see the resonance pole from the threshold, which explains the background, the residue phase, and the difference between the pole and Breit-Wigner parameters. 
\end{abstract}


\maketitle


A resonance is defined as a simple pole of the scattering amplitude in the specific complex-energy region known as the non-physical Riemann sheet \cite{DalitzMoorhouse}. The real and the imaginary part of the pole's position are directly related to resonant mass and width, while the magnitude of the pole's residue gives the branching fraction \cite{PDG}. The physical meaning of the fourth parameter, the residue's phase, is still not clear. For some nucleon resonances this phase seem to depend on the geometric relation between the resonant pole and zeros of the amplitude \cite{Cec17}.

The scattering amplitude is dominated by a resonant pole close to the physical region, i.e., the real-axis of the complex-energy plane above the threshold. It is common to divide such amplitude into two parts: a large resonant term, and a non-resonant background. The resonant part of the amplitude may have resonant parameters in a simple Breit-Wigner \cite{QFT,BW,AtkinsFriedman,NuclPh,Taylor,Boehm,Hoehler,BransdenMoorhouse} or in some more advanced form. Those more advanced forms go from energy-dependent resonant widths \cite{GounarisSakurai,Flatte,ManleySaleski} to highly intricate models that incorporate analyticity and unitarity of the scattering matrix \cite{Cutkosky,Batinic,Sok15}. The resonant parameters are then extracted from the amplitude using analytic continuation and other advanced mathematical methods \cite{Cec08,Sva14,Pel16}. 

Regarding the background, the quantum theory of potential scattering suggests that it stems from the long-range interaction \cite{Boehm}. In excited-nucleon physics \cite{Hoehler}, the background amplitude includes various analytic features of the scattering model such as the left-hand cut and the tails of more distant resonances. It is  sometimes assumed that the background will be small for a resonance sufficiently close to the threshold  (see, e.g., Ref.~\cite{Taylor}). 

In this Letter we use the formalism from Ref.~\cite{Cec17} on the $f_0(500)$ meson and few other nearly-elastic resonances to show exactly the opposite is the case: the closer the resonance is to the threshold, the larger the background. We show that due to the unitarity of scattering amplitude, this large background means the residue's phase will also be large. Finally, we show that this strange non-Breit-Wigner resonance \cite{PDG} is in fact a Breit-Wigner resonance with a somewhat larger Breit-Wigner mass. 

We begin by separating the scattering phase shift into a resonant and a background part
\begin{equation} \label{BreitWignerformula}
    \delta=\delta_\mathrm{R}+\delta_\mathrm{B}.
\end{equation}
Near the resonance, the background $\delta_\mathrm{B}$ is expected to be  constant. The resonant part is given by 
\begin{equation} \label{PoleParameters}
    \tan\delta_\mathrm{R}= \frac{\Gamma/2}{M-E},
\end{equation}
where $M$ is the resonance mass, $\Gamma$ its decay width, while $E$ is the center-of-mass energy. 

To better understand the intricate interplay between the resonant and background parameters, we study an advanced amplitude for resonant scattering from Ref.~\cite{Cec17}
\begin{equation}\label{Amplitude}
    A=x\,e^{i\left(\delta_\mathrm{R}+\delta_\mathrm{B}\right)}\,\sin\left(\delta_\mathrm{R}+\delta_0\right).
\end{equation}
Here $x$ is the branching fraction, and $\delta_\mathrm{0}$ is defined as
\begin{equation} \label{Delta0}
   \tan \delta_0= \frac{\Gamma/2}{E_0-M},
\end{equation}
where $E_0$ is the elastic threshold energy. This choice of $\delta_0$ ensures that the amplitude is zero at the threshold and the residue's phase is given by $\delta_B+\delta_0$.

Here we focus on elastic resonances, those decaying exclusively to one channel, because $f_0(500)$ is such resonance. In that case, the scattering matrix becomes a scalar function, namely $1+2iA$, and the scattering matrix unitarity condition becomes simply 
\begin{equation}\label{Unitarity}
    \mathrm{Im}\,A=|A|^2.
\end{equation} 

This will be satisfied if $x=0$, or if $x=1$ and $\delta_\mathrm{B}=\delta_0$. The last relation is crucial. For elastic resonances close to the threshold, $\delta_B$ will be large and negative. The same will be the case for the residue's phase, which is in this model equal to $2\delta_B$. Since $f_0(500)$ is not too far from the $\pi \pi$ threshold, we would expect that its residue phase is  quite large. But before drawing any firm conclusions, we need to test this simple model against the data.


\begin{figure*}
\includegraphics[width=0.49\textwidth]{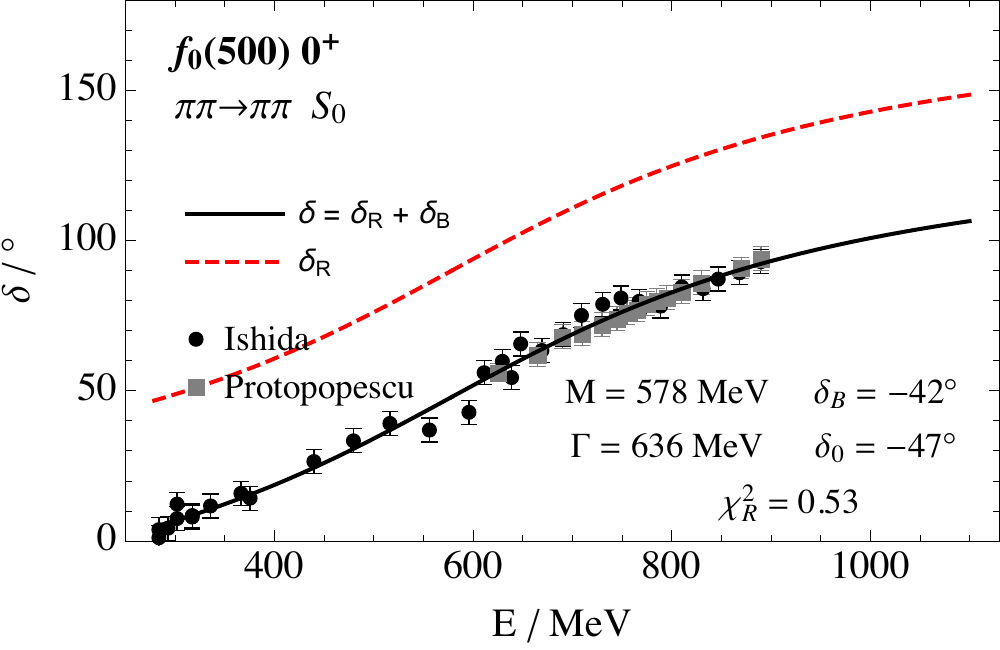}
\includegraphics[width=0.49\textwidth]{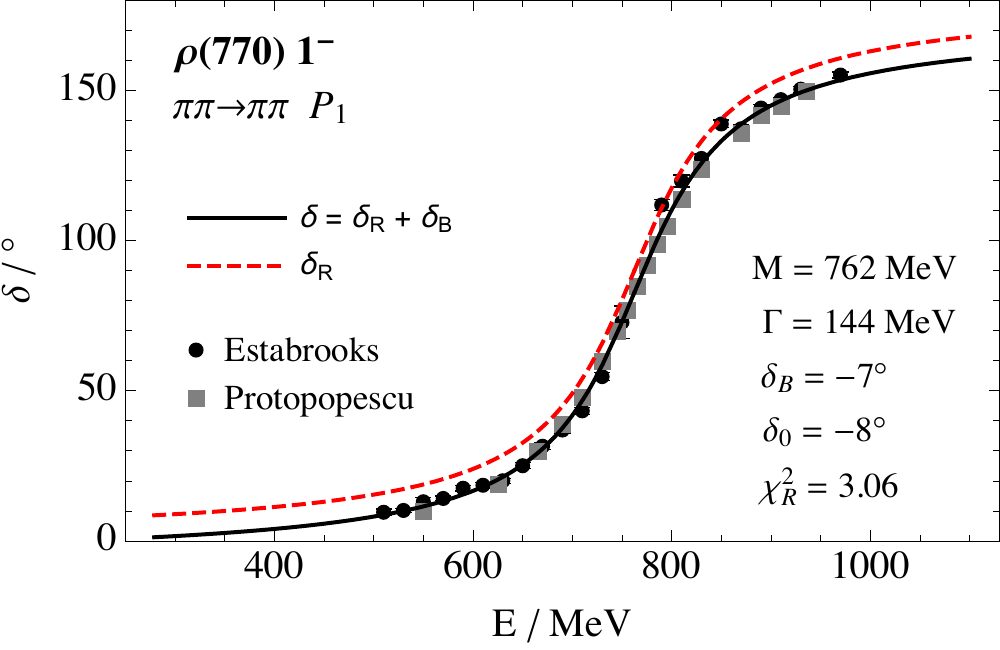}
\includegraphics[width=0.32\textwidth]{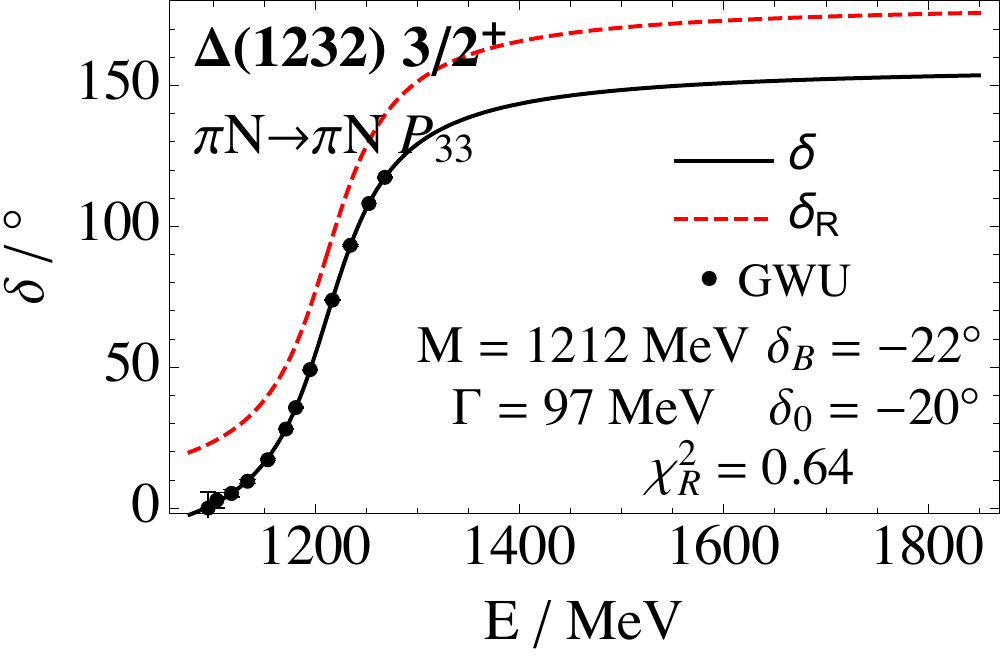}
\includegraphics[width=0.32\textwidth]{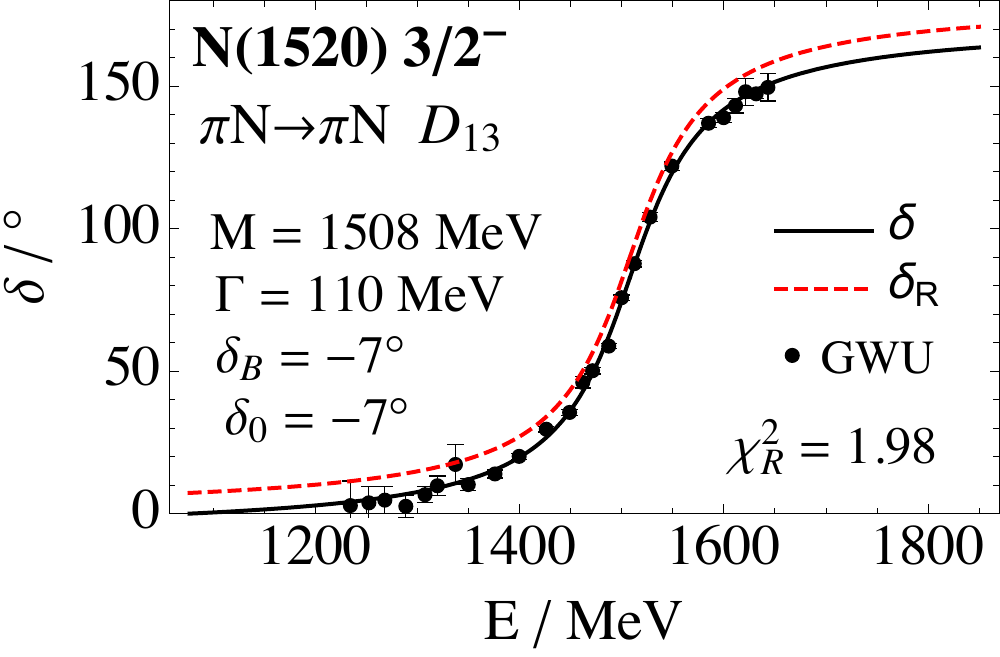}
\includegraphics[width=0.32\textwidth]{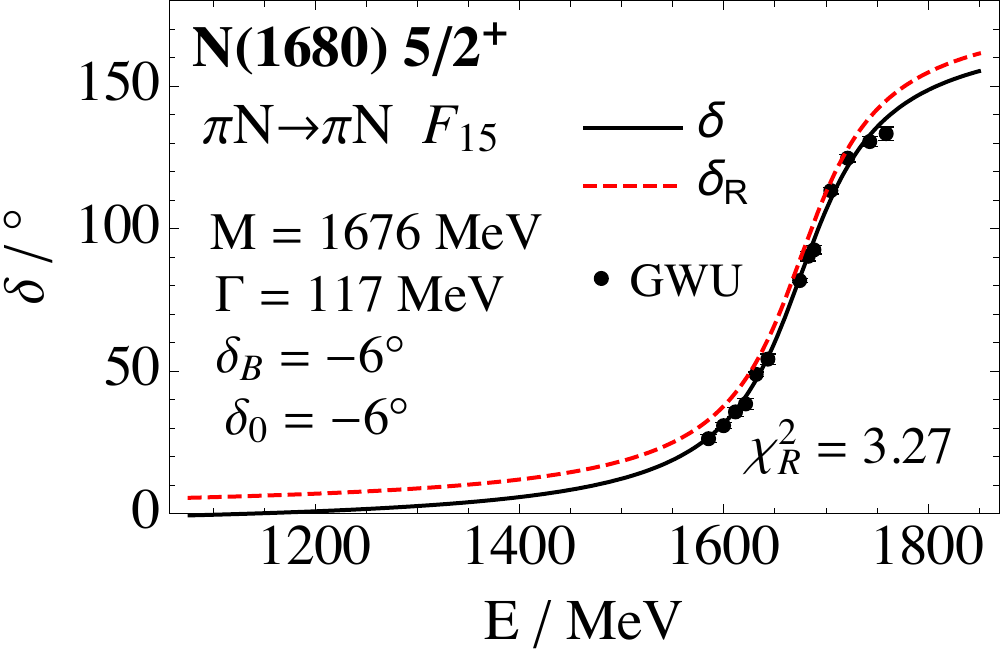}
\caption{Fits of our model to the data. [Upper left graph] The $\pi \pi$ elastic spin-zero isospin-zero phase shifts from Protopopescu \cite{Pro73} (gray squares) and Ishida \cite{Ish97} (black disks). 
[Upper right graph] The $\pi \pi$ spin-one isospin-one elastic phase shifts from Protopopescu \cite{Pro73} (gray squares) and Estabrooks \cite{Est74} (black disks).
[Lower graphs] The phase shifts calculated from the GWU partial-wave single-energy solutions for the $\pi N$ elastic scattering~\cite{Arndt,SAID} 
\label{Fig1}}
\end{figure*}

To see whether $\delta_\mathrm{B}$ is indeed equal to $\delta_0$ for elastic resonances, we fit Eq.~(\ref{BreitWignerformula}) to the available data. The key assumption of this model, that $\delta_\mathrm{B}$ is constant, will be satisfied in the close neighbourhood of the resonance only. But if we are too close, the extracted resonance parameters will be unreliable. We devise the following fitting strategy to bring us to this Goldilocks region. A broad range is chosen around every resonance and all the data points in it are fitted. Then we drop the end-point data that can be identified as belonging to the nearby resonances or in other way drastically depart from the fitted curve. We also demand that the fitted masses and widths are close to the {\it Particle Data Group} (PDG) estimates~\cite{PDG}. 

We are allowed to do all of this because we are not proposing a novel method for the extraction of resonant masses and widths, but trying to get the estimate of a single model parameter, the background phase $\delta_\mathrm{B}$, as good as possible. 

The goodness of the fit is estimated by visual inspection, but also by getting a reasonable value for the reduced chi-squared,  $\chi^2_\mathrm{R}$. This value was seldom in the region expected for the experimental data, slightly above 1, because the phase shifts are not experimental observables. They are not measured directly and their error bars are often no more than rather crude estimates.

First we fit our model to the $\pi \pi$ elastic $S_1$ wave phase shifts from Protopopescu \cite{Pro73} and to a data compilation by Ishida \cite{Ish97}. These are the data where $f_0(500)$ lurks. Since Ishida does not report any errors, we varied error values in the range from $2^\circ$ to $12^\circ$, yet all the fit results turned out to be roughly the same. We have chosen $4^\circ$ for all data points, since this value is used in Protopopescu paper. The fit results are shown in the upper-left part of Fig.~\ref{Fig1}. It  illustrates nicely why it has been extremely difficult even to infer the existence of $f_0(500)$, let alone to determine its properties.

The solid black line is our full fit. The dashed line (red online) shows the resonant part only. The distance between the two is the background  $\delta_\mathrm{B}$. Clearly, the background is really huge for $f_0(500)$. Its fitted value is $-42^\circ$. We compare it to $\delta_0$ calculated using Eq.~(\ref{Delta0}) from the fitted values of $M$, $\Gamma$, and the threshold energy $E_0=279$~MeV, which gives very close result of $-47^\circ$. Our mass of 578~MeV is a bit larger than the PDG estimate (400-550~MeV), but both Ishida and Protopopescu report even larger masses. Our width is 636~MeV, which falls into the PDG region of 400-700~MeV. 

To test our hypothesis against a heavier resonance, we fit the $P_1$ phase shift of the same, $\pi \pi$ elastic scattering amplitude. There we find another elastic resonance, the $\rho(770)$. Results are shown on the right-hand-side of Fig.~\ref{Fig1}. We see that the background of this resonance, which is situated farther from the threshold, is much smaller than the one for $f_0(500)$. Moreover, its value of $-7^\circ$ is almost the same as $\delta_0$'s value of $-8^\circ$.

These were the meson resonances; what about the baryons? To study them, we start with the lowest-mass excited nucleon, the elastic pion-nucleon resonance $\Delta(1232)$. We calculate the phase shifts from the single-energy solutions for the $\pi N$ elastic partial-wave amplitudes done by the {\it George Washington University} (GWU) group \cite{SAID,Arndt}. The background turns out to be rather large. The value of $\delta_\mathrm{B}$ is $-22^\circ$ and the calculated $\delta_0$ differs from the background by just $2^\circ$. The fit is shown in the lower-left part of Fig.~\ref{Fig1}.


There are no other elastic resonances in $\pi N$ scattering, but there are some that decay strongly to the elastic channel. The two most prominent ones are $N(1520)$ and $N(1680)$ with the $\pi N$ branching fractions in the range of 55-65\% and 60-70\% respectively. We tested them against our model as well, and the results make sense. The backgrounds are equal to calculated $\delta_0$s, and become smaller as the resonances are farther from the threshold. The fits are shown in the lower-middle and lower-right parts of Fig.~\ref{Fig1}. 

These results corroborated our main assumption: the background is smaller the farther we are from the threshold. The reason is that the resonant phase shift goes through $90^\circ$ at the resonance mass and the full phase shift is zero at the threshold. It would be interesting to test this finding on similar elastic resonances in other fields of study. But here, we focus on another resonant property that seems to be closely connected to the background phase--the elastic residue's phase $\theta$--which, in our model, is equal to two times the background.

In that sense, we would predict that the elastic residue phase of $f_0(500)$ is rather large and that it has a value of roughly $-84^\circ\pm 4^\circ$, keeping in mind that the error analysis is done on the non experimental data. Another prediction within the same model is that the $\theta$ of $\rho(770)$ is $-15^\circ\pm 1^\circ$. These and all other results are gathered in Table~\ref{Table1} and compared to PDG estimates, where available. To make the comparison easier, we show results for $2\delta_\mathrm{B}$ and $2\delta_0$.

 \begin{table}[h]
\caption{Fit results. Phase $\delta_0$ is calculated using Eq.~(\ref{Delta0}). The elastic residue phase $\theta$ is compared to the $2\delta_\mathrm{B}$    \label{Table1} }
\begin{tabular}{|l|ccc|c|} 
\hline
\hline
Resonance & $M$ & $\Gamma$ & $2\delta_\mathrm{B}$ & $2\delta_0$  \\
 & (MeV) & (MeV) & ($^\circ$) & ($^\circ$)   \\
 \hline
 \hline
 $f_0(500)\,0^+\,[\pi\pi\,S_0]$ & $578\pm 15$ & $636\pm12$ & $-84\pm4$ & $-93\pm4$   \\
 PDG        & $400$-$550$ & $400$-$700$  & N/A &  \\
\hline 
 $\rho(770)\,1^-\,[\pi\pi\,P_1]$ & $762\pm1$ & $144\pm2$ & $-15\pm1$ & $-17\pm1$   \\
 PDG          & $763 \pm 1$ & $150\pm1$ & N/A  &  \\ 
\hline 
  $\Delta(1232)\,\frac{3}{2}^+\,[\pi N\,P_{33}]$ & $1212\pm 1$ & $97\pm1$ & $-44\pm1$ & $-40\pm1$   \\
 PDG        & $1210\pm1$ & $100\pm2$ &$-46\pm^{1}_{2}$ &  \\
 \hline 
 $N(1520)\,\frac{3}{2}^-\,[\pi N\,D_{13}]$ & $1508\pm 1$ & $110\pm2$ & $-15\pm2$ & $-15\pm2$   \\
 PDG         & $1510\pm5$ & $110\pm^{10}_{5}$ & $-10\pm5$&  \\
 \hline 
  $N(1680)\,\frac{5}{2}^+\,[\pi N\,F_{15}]$ & $1676\pm 3$ & $117\pm4$ & $-12\pm4$ & $-11\pm4$   \\
 PDG       & $1675\pm^5_{10}$ & $120\pm^{15}_{10}$ & $-5\pm15$&  \\
\hline\hline
\end{tabular}
\end{table}

We must note here that the PDG estimates of $\theta$ may be tricky, because the residue phase is notoriously hard to extract properly. To make more sense out of it,  we explicitly write the phases from the most important analyses here. For $N(1520)$, the values of $\theta$ that PDG uses for the average are $-12^\circ\pm5^\circ$ \cite{Cutkosky}, $-15^\circ\pm2^\circ$ \cite{Sva14}, and $-14^\circ\pm3^\circ$ \cite{Sok15}. We see that our result of $-15^\circ\pm2$ fits perfectly. For $N(1680)$, the PDG values are widely scattered: $-25^\circ\pm5^\circ$ \cite{Cutkosky}, $-16^\circ\pm2^\circ$ \cite{Sva14}, and $+5^\circ\pm10^\circ$ \cite{Sok15}. Our model could be of help in such situations, since our result of $-12^\circ\pm4^\circ$ points towards exactly one of those phases.


We find useful to show a combined plot of all our results to illustrate the main finding of this Letter: the geometrical relation between the background, and resonant masses and widths. The $x$-axis of this plot shows how far the resonant mass $M$ is from the threshold. In this way we can show both meson and baryon resonances on the same plot. The $y$-axis shows the resonance half-width $\Gamma/2$. For each resonance we put "*" at the position of its fitted values. To check the self-consistency of our model, we use the fitted mass and  background phase and calculate what would be the half-width value assuming fitted $\delta_\mathrm{B}$ is $\delta_0$. For this, we use the inverse of Eq.~(\ref{Delta0}). Then we plot "+" at the position of the fitted mass and the newly calculated half-width. If our model makes sense, the plus sign and the asterisk should be reasonably close. 

\begin{figure}[h]
\includegraphics[width=0.47\textwidth]{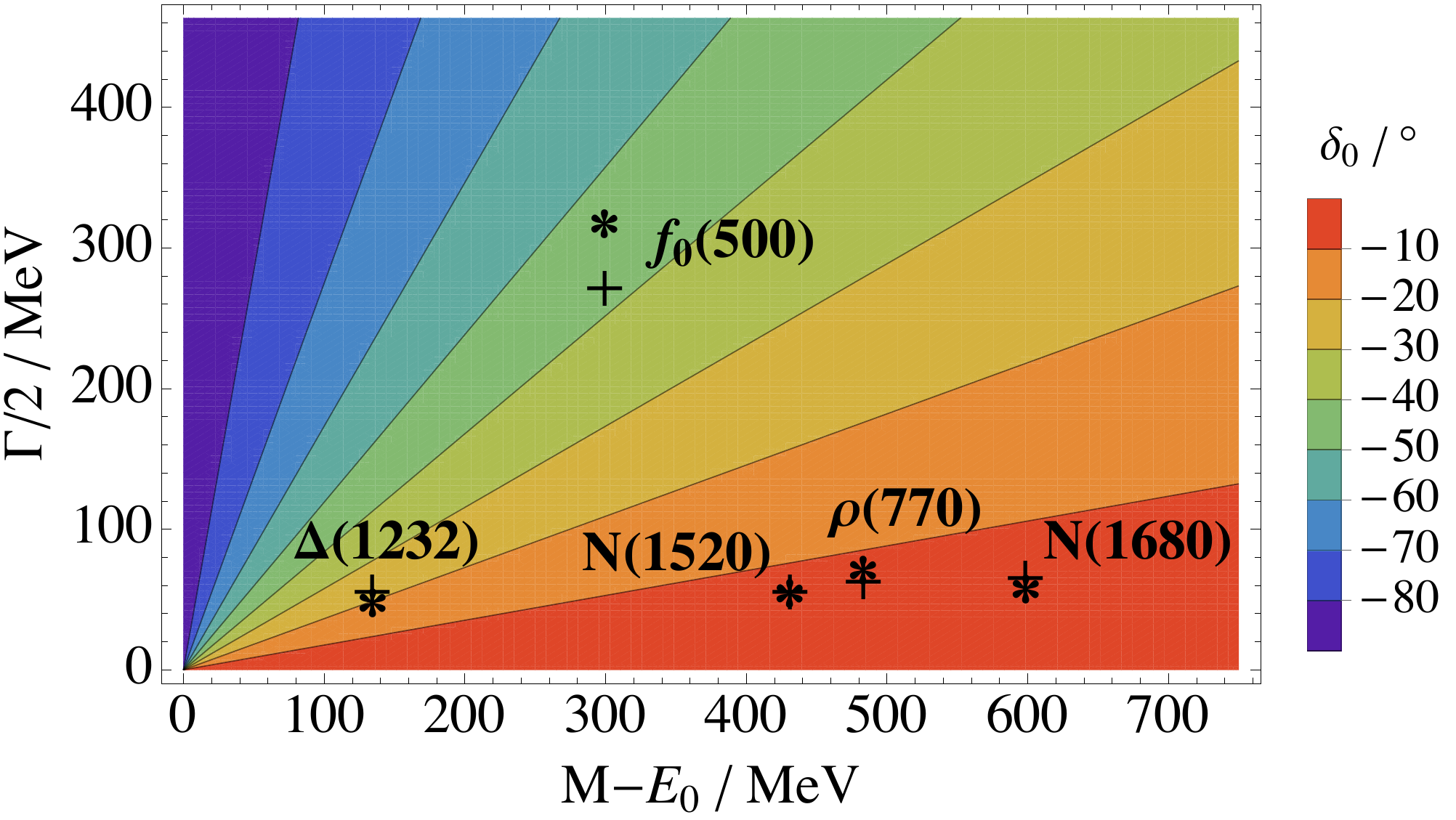}
\caption{The self-consistency plot. Each ``*" is at the position of the fitted $M$ and $\Gamma$, while for each ``+" we calculate $\Gamma$ using Eq.~(\ref{Delta0}) from fitted values of $\delta_\mathrm{B}$ and $M$ (with $\delta_0=\delta_\mathrm{B}$). 
\label{Fig2}}
\end{figure}

As can be seen from Fig.~\ref{Fig2}, that is true even for $f_0(500)$, while for all other resonances the two symbols are almost indistinguishable.

Before we conclude, one more thing needs to be clarified. The formula given in Eq.~(\ref{BreitWignerformula}) is often used in the literature~\cite{Taylor,Boehm,Hoehler}. However, some researchers use a somewhat different version~\cite{BransdenMoorhouse}
\begin{equation}
\label{BreitWignerAlternative}
\tan\widetilde{\delta}=\tan\widetilde{\delta}_\mathrm{R}\,+\tan\widetilde{\delta}_\mathrm{B},
\end{equation}
with the resonant phase given by
\begin{equation}
    \tan\widetilde{\delta}_\mathrm{R}= \frac{\widetilde{\Gamma}/2}{\widetilde{M}-E}.
\end{equation}
If we assume that the background phase $\widetilde{\delta}_\mathrm{B}$ is constant, the $\widetilde{\delta}$ and $\delta$ from Eq.~(\ref{BreitWignerformula}) will have exactly the same mathematical form. The mathematical equivalence between the two phase shifts means that there is no need for new fits. Moreover, it turns out that $\widetilde{\delta}_\mathrm{B}=\delta_\mathrm{B}$. Yet, since the sum of tangents is not equal to the tangent of the sum, resonant parameters extracted with this formula will have different values from those obtained by Eq.~(\ref{PoleParameters}). These alternative parameters are given by 
\begin{eqnarray}
\widetilde{M}&=& M-(\Gamma/2)\,\tan\delta_\mathrm{B},\label{BWM}\\   \widetilde{\Gamma}&=&\Gamma/\cos^2\delta_\mathrm{B}\label{BWGamma}.
\end{eqnarray}

One does not simply compare these parameters to the PDG resonant parameter from Table~\ref{Table1}. There, we compared masses and widths to the real and imaginary parts of the resonant pole. However, in addition to the PDG tables with pole parameters, some resonances have also tables with the so called Breit-Wigner parameters, $M_\mathrm{BW}$ and $\Gamma_\mathrm{BW}$. Moreover, Eqs.~(\ref{BWM}) and (\ref{BWGamma}) are given exactly as the formulas for the Breit-Wigner mass and width in Ref.~\cite{Cec17}. Therefore, we calculate the values of these alternative (tilde) parameters using the results of our previous fit and compare them to the PDG estimates for their Breit-Wigner parameters, where available.

The big question here is, do we really need the Breit-Wigner mass and width at all. We could simply stop here to avoid producing further confusion with all these additional parameters.  We would have agreed that it is extremely reasonable to choose one set of resonant parameters and stick to it, had we not studied the Z boson. 

In the case of the Z boson, PDG reports the Breit-Wigner parameters only. These parameters are then used to build the scattering amplitude in PDG \cite{PDG} as $A_{if}=x_{if}\,A_Z$, where $x_{if}$ is the geometrical average of the branching fractions for the Z boson decaying to initial and final channels of this scattering, and
\begin{equation}
\label{Zboson}
A_Z=\frac{\Gamma_\mathrm{BW}\,E^2/M_\mathrm{BW}}{M_\mathrm{BW}^2-E^2-i\,\Gamma_\mathrm{BW}\,E^2/M_\mathrm{BW}}.
\end{equation}

Eq.~(\ref{Zboson}) appears to be quite different from our simple model; these BW parameters should not have anything to do with our alternative parameters. Yet, since $A_Z$ satisfies the unitarity constraint in  Eq.~(\ref{Unitarity}), we apply our formalism to it. (Note also that the $A_{if}$ automatically satisfies the matrix unitarity constraint $\mathrm{Im}\,A=A^\dag A$.) 

The Breit-Wigner mass of the Z boson is 91~188~MeV and the width is 2495~MeV \cite{PDG}. We calculate the pole position using Eq.~(\ref{Zboson}). The real part of the pole, the mass $M$, is 91~165~MeV. The imaginary part multiplied by $-2$ gives the width $\Gamma$, which turns out to be 2494~MeV. Since the width of the Z boson is several times larger than the width of $f_0(500)$, one  might in light of this Letter conclude that the residue phase would be very large. However, the distance to the threshold is numerous times larger then the width, and therefore the residue phase should, in fact, be tiny. And indeed, by calculating the residue we obtain the phase of $-2.4^\circ\pm0.2^\circ$. 

Assuming that this residue phase is twice the background phase $\delta_\mathrm{B}$, as it is in our model, we calculate the alternative parameters $\widetilde{M}$ and $\widetilde{\Gamma}$ for the Z boson. With known $M$, $\Gamma$, and $\delta_\mathrm{B}$, we simply use Eqs. (\ref{BWM}) and (\ref{BWGamma}). For $\widetilde{M}$ we get 91~188~MeV, while $\widetilde{\Gamma}$ is 2495~MeV. Our alternative parameters turn out to be exactly the Breit-Wigner parameters we began with. And that is why we choose to show here the Breit-Wigner parameters as well. 

All the parameters in this Letter seem to be different parts of the same story. The story about the unitary scattering matrix which can, close to the resonance, be approximated by a simple pole plus constant background. Because of the simplicity, there will be just a few parameters, and because of the unitarity, all of them cannot be independent. And that is what we have found.

We calculate the alternative parameters for all of the resonances we have considered here and show them in Table \ref{Table2}, where we compare them to the known PDG Breit-Wigner parameters \cite{PDG}. Since for the $\rho$ meson there is no PDG estimate, we use the  Breit-Wigner mass and width from Ref.~\cite{Hus86}. 

\begin{table}[h]
\caption{ Alternative vs.~the Breit-Wigner parameters (PDG estimates \cite{PDG}, and Ref.~\cite{Hus86} for the $\rho$ meson) \label{Table2} }
\begin{tabular}{|l|cc|cc|} 
\hline\hline 
Resonance & $\widetilde{M}$ & $M_\mathrm{BW}$ & $\widetilde{\Gamma}$ & $\Gamma_\mathrm{BW}$ \\
 & (MeV) & (MeV) & (MeV) & (MeV)   \\
 \hline\hline
 $f_0(500)$ & $866\pm 8$  & $400$-$550$& $1156\pm78$ & $400$-$700$   \\
 $\rho(770)$ & $772\pm1$  & $771\pm 4$ & $147\pm2$ & $150\pm 5$ \\ 
 \hline
  $\Delta(1232)$ & $1231\pm 1$ & $1232\pm2$ & $113\pm1$ & $117\pm3$  \\
 $N(1520)$ & $1515\pm 1$ & $1515\pm5$ & $112\pm3$ & $110\pm10$  \\
  $N(1680)$ & $1682\pm 1$ & $1685\pm5$ & $119\pm4$ & $120\pm^{10}_{5}$ \\ 
  \hline
  $Z$ boson & $91\,188\pm2$  & {$91\,188 \pm 2$} & $2495\pm2$  & {$2495\pm2$} \\
\hline\hline
\end{tabular}
\end{table}

In the table, everything fits almost perfectly, apart from the $f_0(500)$. That is not unexpected because it is often stated that $f_0(500)$ certainly cannot be modeled as a naive Breit-Wigner resonance \cite{PDG}. Most of those advanced models try to put the value of the Breit-Wigner mass close to the real part of the pole. Yet, we show here that if $f_0(500)$ is allowed to have a larger Breit-Wigner mass due to the larger background phase, which is large because the real part of the pole is close and imaginary part far from the threshold, $f_0(500)$ would not be different from other  Breit-Wigner resonances at all.

\begin{acknowledgments}
S.C.~would like to thank Lothar Tiator, Hedim Osmanovi\' c, Jovana Nikolov, and Kre\v simir Jakov\v ci\' c for invaluable discussions, comments, and suggestions.
\end{acknowledgments}

\end{document}